\documentclass[10pt, conference]{IEEEtran}

\usepackage[T1]{fontenc}
\usepackage[latin9]{inputenc}
\usepackage{geometry}
\usepackage{amsmath}
\usepackage{xpatch}
\usepackage{amssymb}
\usepackage{esint}
\usepackage{mathtools}
\usepackage{amsthm}
\usepackage{nicefrac}
\usepackage{bigints}
\usepackage{blkarray, bigstrut}
\usepackage{physics}
\usepackage{calligra}
\usepackage{graphicx}
\usepackage{epstopdf}
\usepackage{dsfont}
\usepackage{array,ragged2e}
\usepackage{enumitem}
\usepackage{etoolbox}
\usepackage[english]{babel} 

\usepackage[nopar]{lipsum}
\usepackage{psfrag}
\usepackage[ruled,lined,linesnumbered]{algorithm2e}
\usepackage{algorithmic}
\usepackage{algorithm2e}
\usepackage{balance}
\usepackage{float}
\usepackage{hyperref}
\usepackage{subcaption}
\usepackage{soul}

\SetKw{KwBy}{by}

\makeatletter
\newcommand{\removelatexerror}{\let\@latex@error\@gobble}
\makeatother

\graphicspath{ {Figures/} }

\xpatchcmd{\proof}{\hskip\labelsep}{\hskip5\labelsep}{}{}  
\makeatletter
\xpatchcmd{\proof}{\@addpunct{.}}{\@addpunct{:}}{}{}
\makeatother

\renewcommand\[{\begin{equation}}
\renewcommand\]{\end{equation}} 
\usepackage{listings}
\usepackage{fancyvrb}
\usepackage{framed}

\usepackage{courier}
\usepackage[usenames,dvipsnames,table]{xcolor}

\definecolor{dkgreen}{rgb}{0,0.3,0}
\definecolor{gray}{rgb}{0.5,0.5,0.5}

\makeatletter
\newcommand*{\rom}[1]{\expandafter\@slowromancap\romannumeral #1@}
\makeatother

\newlength{\oldtextfloatsep}
\usepackage{siunitx}
\usepackage{tabu}
\usepackage{booktabs}
\usepackage{multirow}
\usepackage{capt-of}
\usepackage{array}
\usepackage{arydshln}
\newcommand{\comment}[1]{}

\usepackage{url}
\makeatletter
\patchcmd{\@maketitle}
  {\addvspace{0.5\baselineskip}\egroup}
  {\addvspace{-1\baselineskip}\egroup}
  {}
  {}
\makeatother

\begin{document}

\title{

X-GRL: An Empirical Assessment of Explainable GNN-DRL in B5G/6G Networks

}

\author{\IEEEauthorblockN{
Farhad Rezazadeh\IEEEauthorrefmark{1}, Sergio Barrachina-Mu\~noz\IEEEauthorrefmark{1}, Engin Zeydan\IEEEauthorrefmark{1},\\
Houbing Song\IEEEauthorrefmark{3},
K.P. Subbalakshmi\IEEEauthorrefmark{4},
and Josep Mangues-Bafalluy\IEEEauthorrefmark{1}
}

\IEEEauthorrefmark{1}\normalsize{}Centre Tecnol\'ogic de Telecomunicacions de Catalunya (CTTC), Barcelona, Spain\\

\IEEEauthorrefmark{3}University of Maryland, Baltimore County (UMBC), Baltimore, USA\\
\IEEEauthorrefmark{4}Stevens Institute of Technology, New Jersey, USA\\

{\normalsize{}Contact Emails:  \texttt{\{name.surname\}@cttc.es},      ~\texttt{h.song@ieee.org},~\texttt{	ksubbala@stevens.edu}
}
}

\maketitle

\begin{abstract}

The rapid development of artificial intelligence (AI) techniques has triggered a revolution in beyond fifth-generation (B5G) and upcoming sixth-generation (6G) mobile networks. Despite these advances, efficient resource allocation in dynamic and complex networks remains a major challenge. This paper presents an experimental implementation of deep reinforcement learning (DRL) enhanced with graph neural networks (GNNs) on a real 5G testbed. The method addresses the explainability of GNNs by evaluating the importance of each edge in determining the model's output. The custom sampling functions feed the data into the proposed GNN-driven Monte Carlo policy gradient (REINFORCE) agent to optimize the gNodeB (gNB) radio resources according to the specific traffic demands. The demo demonstrates real-time visualization of network parameters and superior performance compared to benchmarks.

\end{abstract}

\section{Introduction}

\IEEEPARstart{6}{G} wireless communication systems herald a new era of high-speed, low-latency, reliable connectivity. They inherently support AI to pave the way for new services and use cases. Nevertheless, the opaque nature of AI can undermine trustworthiness and hinder its widespread use in critical applications. Consequently, there is a growing need for explainable AI (XAI)\cite{XAI-Song}\cite{XAI-Farhad}\cite{XAI-Subbalakshmi}, whose primary goal is to shed light on decision-making processes. This paper investigates the feasibility of integrating Graph Convolutional Networks (GCNs) into the REINFORCE~\cite{REINFORCE-Zhang} algorithm in a 5G network to control the allocation of physical resource blocks (PRBs).

\begin{figure}[t!]
\centering
\includegraphics[width=.9\columnwidth]{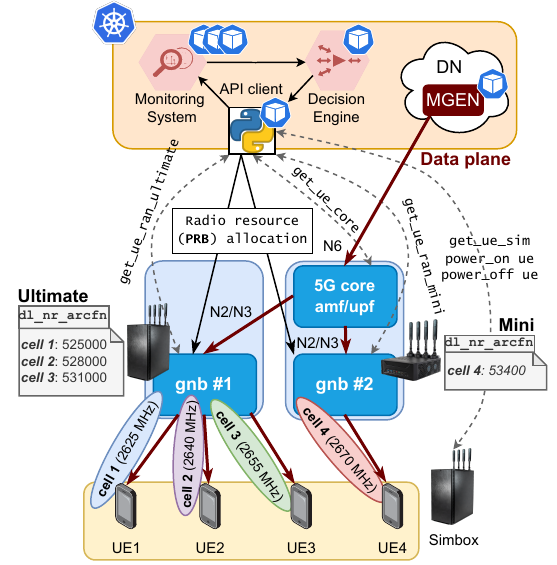}   
\caption{Infrastructure and network setup.}
\label{fig:testbed_architecture}
\end{figure}

The main idea is that while the original state data from the network environment is not inherently structured or graph-like, our solution constructs a graph structure from this state data to enable further processing with GCNs. In our model, we conceptualize the states of the 5G network as a graph in which each state represents a node. Our main focus is to explore the graph convolution operation, a fundamental component of GCNs. In short, graph convolution is about updating node features using its characteristics and the features of neighboring nodes. This process facilitates the distribution of information across the nodes of the graph and creates a form of \emph{communication} between them. More precisely, the operation of graph convolution takes place on the input features \emph{(x)} and the edge index (edge-index) in the forward propagation process. During this process, the attributes of each node are updated taking into account its neighbors, and the resulting updated node features are stored in the variable \emph{x}. Furthermore, the importance of the edge in the graph is elucidated using the proposed explainer function that adopts an optimization-based approach. The edge mask obtained by this function provides valuable insight and explainability into the model's decision process for particular nodes in the graph.

\section{Testbed Architecture}
\begin{figure}[t!]
	\centering
	\includegraphics[width=0.32\textwidth] {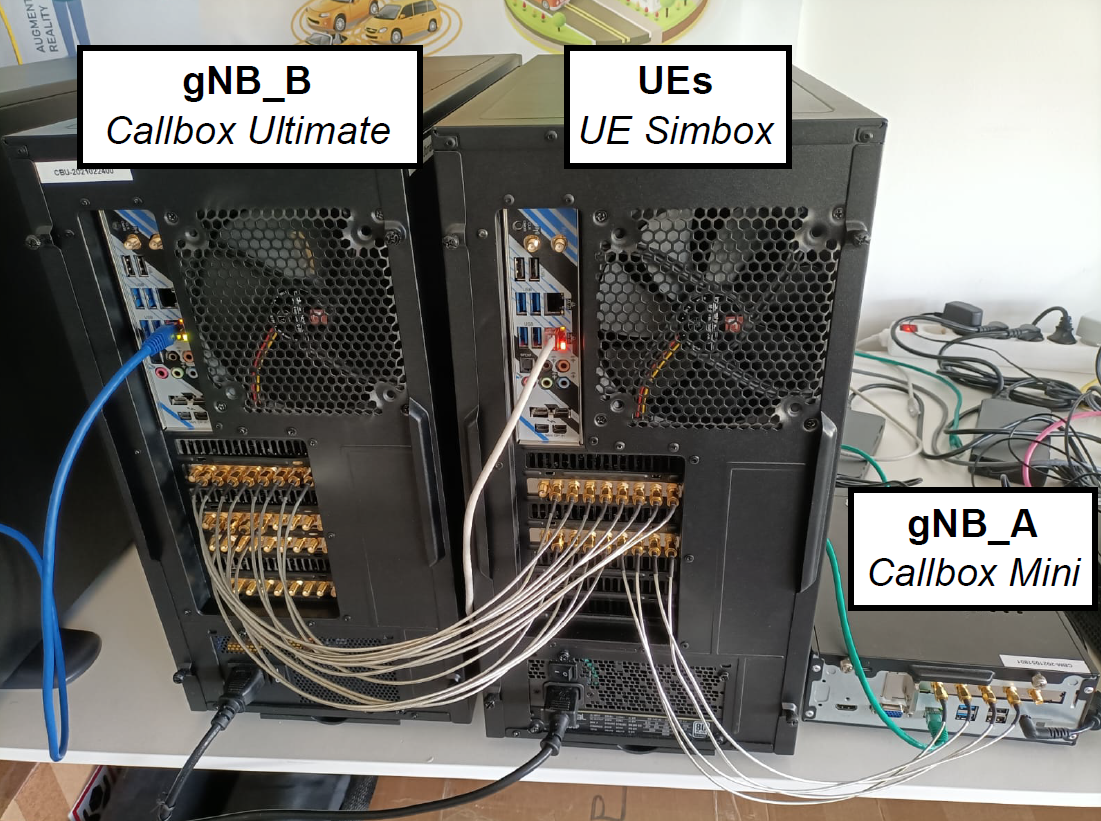}
	\caption{5G RAN components.}\label{fig:amarisoft}
\end{figure}

The testbed is shown in Fig.~\ref{fig:testbed_architecture} and consists of specific hardware and software components to realize the 5G infrastructure and management-related functionalities. For the implementation of the gNBs and the 5G core, the testbed uses a Callbox Ultimate and a Callbox Mini, respectively. To emulate user equipment (UE) interactions with the 5G network, the testbed incorporates a Simbox. These Amarisoft components were selected for their ability to accurately replicate the behavior of 5G elements in a controlled environment. The physical setup of RAN with the Amarisoft devices is shown in Fig.~\ref{fig:amarisoft}. 

On top of this 5G infrastructure, the testbed deploys a cloud-native platform. This Kubernetes-based platform hosts three main management-related workloads. First, the API client processes requests to the 5G network elements. Second, the monitoring system continuously collects relevant metrics from all components of the system to help with real-time performance analysis and system evaluation. This monitoring system is implemented as a collection of pods that contain a Kafka bus along with custom sampling functions that span multiple domains. And finally, the decision engine performs actions at the RAN level. As for the data plane, the testbed uses MGEN~\footnote{\url{https://github.com/USNavalResearchLaboratory/mgen}}\cite{rezazadeh2023multi} to generate downlink traffic patterns of interest.


\begin{figure}[ht]
\centering
\includegraphics[width=.7\columnwidth, clip,trim={0cm 0cm 0cm 0cm}]{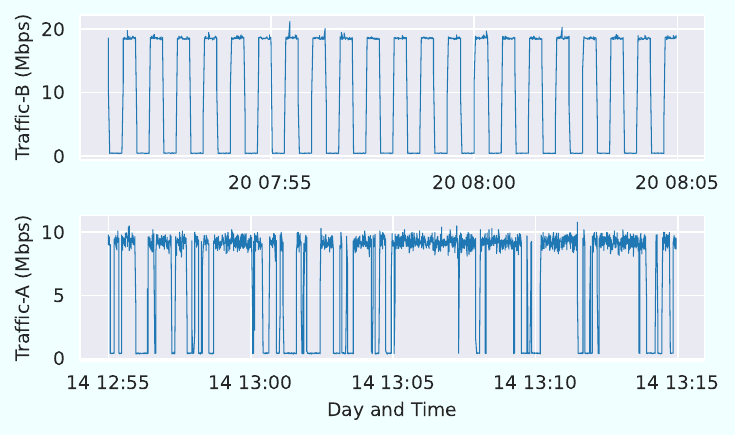}
\caption{Traffic patterns generated throughout the demo.}
\label{fig:traffic}
\vspace{-.4cm}
\end{figure}

\begin{figure}[t!]
\centering
\includegraphics[width=.7\columnwidth, clip,trim={0cm 0cm 0cm 0cm}]{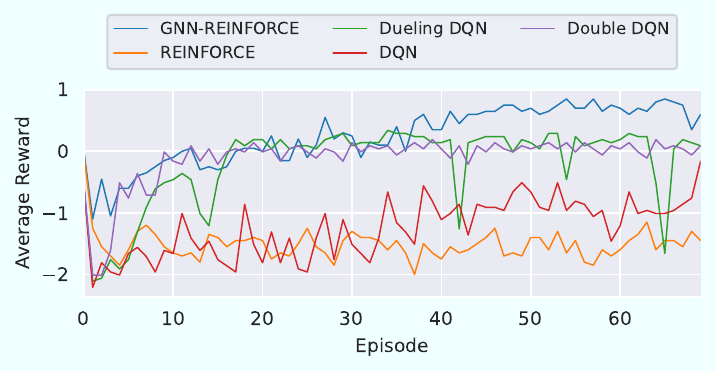}
\caption{Performance comparison of various DRL approaches for radio resource allocation in the training phase.}
\label{fig:Reward}

\end{figure}

\section{Operational Phases}

As shown in Fig.~\ref{fig:traffic}, we examine two different traffic patterns, which are then forwarded to Simbox as downlink traffic. The pattern \emph{A} has a stochastic character. In this case, the packets arrive according to a Poisson process, where the arrival time between packets corresponds to an exponential distribution. This means that packets can arrive at any time, regardless of when the previous packet was received. On the other hand, the traffic pattern \emph{B} adheres to a periodic pattern and transmits data packets at regular intervals. This regularity facilitates learning because of the predictable data rate.

\begin{figure*}[t]
    \centering
    \begin{subfigure}[b]{0.3\textwidth}
        \includegraphics[width=\textwidth]{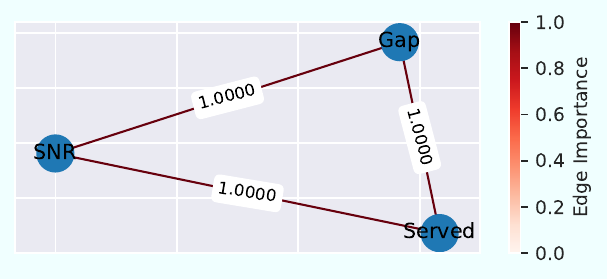}
        \caption{Early training stage}
        \label{fig:1}
    \end{subfigure}
    \hfill
    \begin{subfigure}[b]{0.3\textwidth}
        \includegraphics[width=\textwidth]{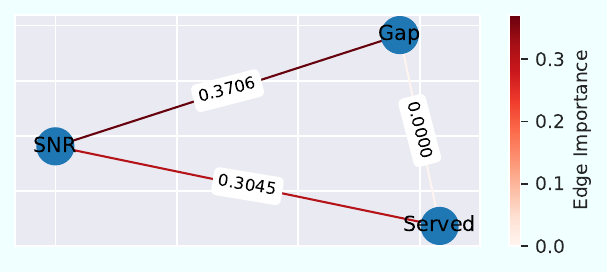}
        \caption{Mid-Training stage}
        \label{fig:2}
    \end{subfigure}
    \hfill
    \begin{subfigure}[b]{0.3\textwidth}
        \includegraphics[width=\textwidth]{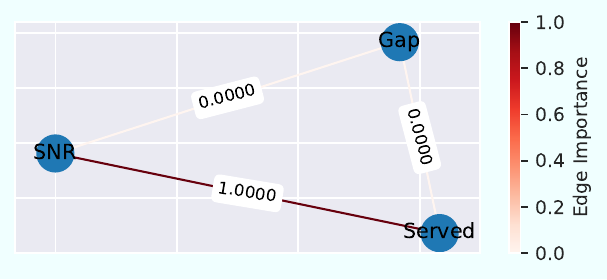}
        \caption{Post-Training analysis}
        \label{fig:3}
    \end{subfigure}
    \caption{Explainability analysis of the GNN-REINFORCE algorithm. }
    \label{fig:XAI}
    \vspace{-.1cm}
\end{figure*}

The decision engine, developed in Python, uses the OpenAI Gym\footnote{\url{https://gymnasium.farama.org/}}\cite{Specialization_TVT} toolkit to interface with DRL agents. Fig.~\ref{fig:Reward} illustrates the average reward per episode for five different algorithms with a single UE. The reward function indicates the effectiveness of decision making, with higher rewards indicating a more optimal allocation method. Our proposed GNN-REINFORCE algorithm ensures superior learning generalization and robust performance compared to other DRL approaches. At the beginning of the training, the agent starts exploring the action space, i.e. the PRBs chunks. This concept of exploration is ingrained through the probabilistic aspect of action selection, a fundamental feature of policy gradient techniques. The agent then tries to find an optimal balance between the learned decision policies and the network states, indicating the exploitation phase. The inclusion of a graph convolution operation in the GNN-REINFORCE algorithm significantly improves the performance of the agent compared to the standard REINFORCE algorithm.

Fig.~\ref{fig:XAI} gives an insight into the explainability of the GNN-REINFORCE algorithm by highlighting the importance of edge connections. This analysis explores the dynamic evolution of edge importance values over time within the GNN. It sheds light on the learning process of the network and how it gradually prioritizes edges between nodes. During the initial training phase, we observe that all edges have an equal importance value of \emph{one}. This indicates the algorithm is starting to learn the relationships between the nodes. We notice a variation in the edge importance values as training progresses to the mid-training phase. The weights are starting to be adjusted on the feedback received from the network environment. In the post-training or inference phase, a transformation is evident. Only one edge is marked as important, with a non-zero value, while the importance values of all other edges have diminished to \emph{zero}. This evolution signifies the key edge (or relationship) that is most influential in making decisions or predictions.

\begin{figure}[ht]
    \centering
 
    \begin{subfigure}[b]{0.35\textwidth}
        \includegraphics[width=\textwidth]{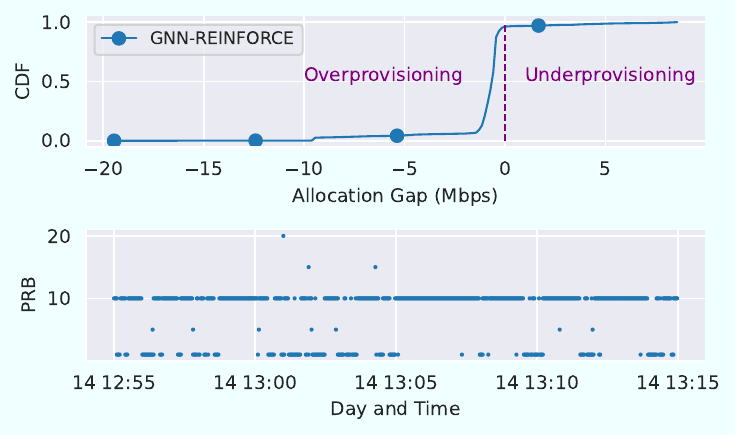}
        \caption{Under Traffic-A.}
        \label{fig:22}
    \end{subfigure}
       \begin{subfigure}[b]{0.35\textwidth}
        \includegraphics[width=\textwidth]{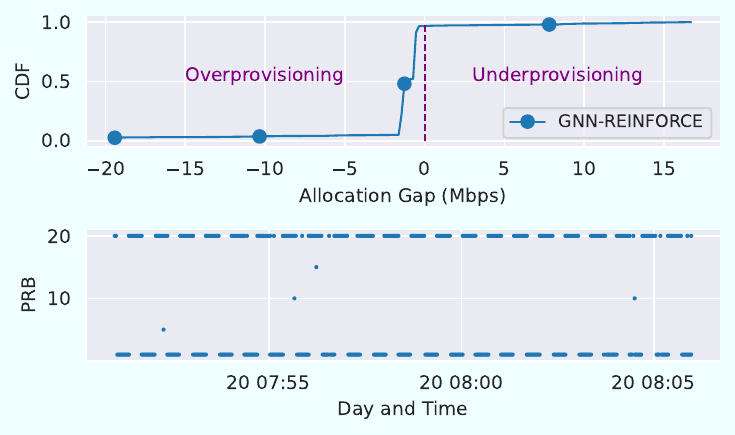}
        \caption{Under Traffic-B.}
        \label{fig:11}
    \end{subfigure}
    \caption{Analysis of network performance under Traffic A and B. (Up) The  CDF of radio resource allocation gap, (Down) Performance evaluation in terms of PRB allocation.}
    \label{fig:P-TRAFFIC}
    \vspace{-.09cm}
\end{figure}

Fig.~\ref{fig:P-TRAFFIC}-(a) and (b) show the network performance of the GNN-REINFORCE algorithm under Traffic-A and Traffic-B, respectively. Each plot is divided into two main sections showing the cumulative distribution function (CDF) of the gap in the allocation of radio resources and the agent's performance in the allocation of PRBs. In 5G network management, the radio resource allocation gap is typically defined as the discrepancy between the expected or desired allocation of a radio resource and the actual allocation, which directly affects the overall performance of the network.

The CDF provides a statistical assessment of the efficiency of resource allocation. Zero gap values imply ideal allocation, indicating more effective use of resources. The performance of the GNN-REINFORCE agent demonstrates a remarkable ability to allocate PRBs. It allocates resources according to traffic demand with a remarkable accuracy of more than \emph{90\%}. 

\begin{figure}[ht]
\centering
\includegraphics[width=.7\columnwidth, clip,trim={0cm 0cm 0cm 0cm}]{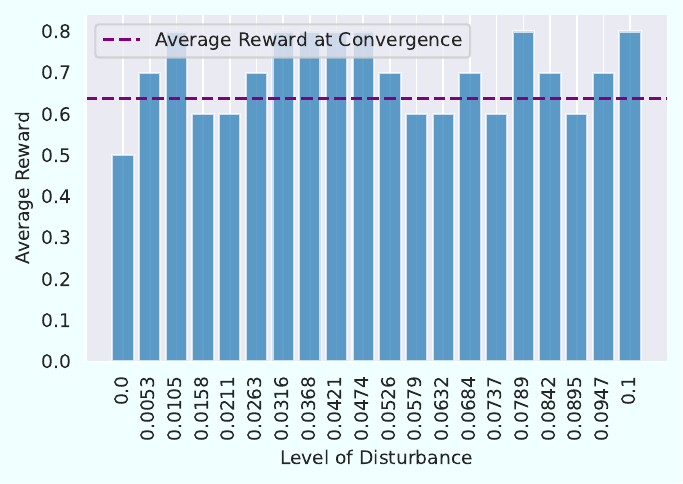}
\caption{Robustness of the model in inference phase.}
\label{fig:Robustness}
\end{figure}

\begin{figure}[ht]
\centering
\includegraphics[width=.7\columnwidth, clip,trim={0cm 0cm 0cm 0cm}]{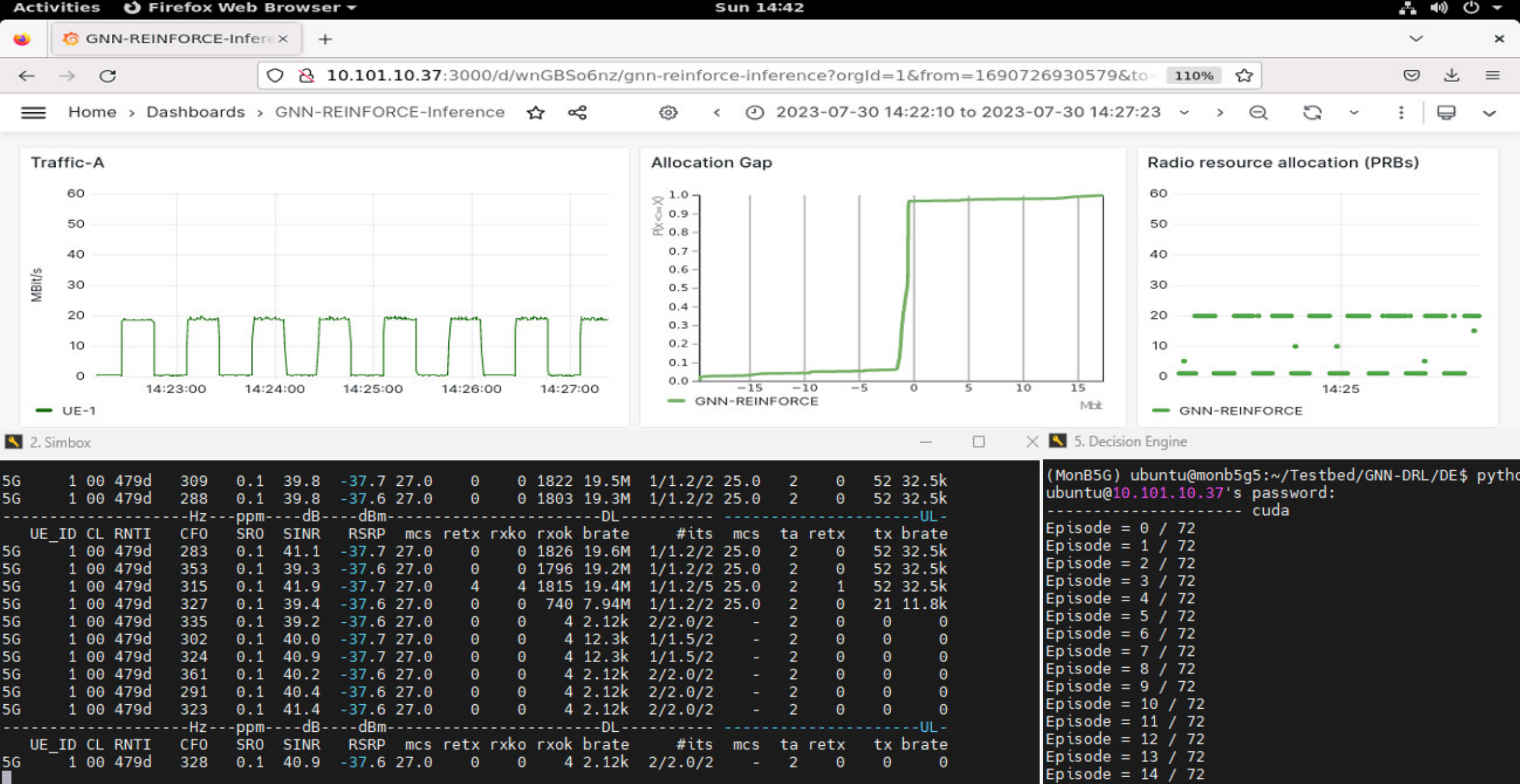}
\caption{Visualization of metrics of the inference phase with Grafana.}
\label{fig:garafana}
\vspace{-.5cm}
\end{figure}

In Fig.~\ref{fig:Robustness}, we evaluated the robustness of our GNN-REINFORCE model by introducing Gaussian noise into the input state values, with noise values ranging from 0 to 0.1. We established these disturbance levels by creating 20 numbers evenly distributed in this range, each representing the standard deviation of the Gaussian noise. We then exposed the model to each disturbance level and logged its performance. The objective of this evaluation was to investigate the ability of the model to maintain its performance at various noise disturbances. As shown in Fig.~\ref{fig:Robustness}, the model showed satisfactory performance in the inference phase compared to the average reward at convergence. Fig.~\ref{fig:garafana} shows the inference phase and the main metrics monitored and observed on Grafana dashboard.

\section{Conclusion}
The synergy of DRL, GNNs, and XAI offers a compelling path towards enhanced efficiency, transparency, and reliability in mobile network management.  Future work will extend this approach to larger and more complex network scenarios and explore potential applications in other areas of network operations and management.
\section*{Acknowledgment}
This work was partially funded by MCIN/AEI/ 10.13039/501100011033 grant PID2021-126431OB-I00 (ANEMONE), Spanish MINECO grant TSI-063000-2021-54 (6G-DAWN) and grant TSI-063000-2021-56 (6G-BLUR), Generalitat de Catalunya grant 2021 SGR 00770 (6GE2E),

\end{document}